\documentclass[noshowpacs,amsmath,twocolumn,superscriptaddress,aps,prl,groupedaddress]{revtex4-1}

\usepackage{setspace}
\usepackage{amsmath}
\usepackage{graphicx}
\usepackage{verbatim}  
\usepackage{amsfonts}
\usepackage{amssymb}
\usepackage{epstopdf} 
\usepackage{xcolor}
\DeclareGraphicsExtensions{.pdf,.eps,.png,.jpg,.mps} 

\begin{document}

\title{The Phonon-Limited-Linewidth of Brillouin Lasers at Cryogenic Temperatures }


\author{Myoung-Gyun Suh, Qi-Fan Yang, and Kerry J. Vahala}
\email{vahala@caltech.edu}
\affiliation{T. J. Watson Laboratory of Applied Physics, California Institute of Technology, Pasadena, California 91125, USA.}

\begin{abstract}
Laser linewidth is of central importance in spectroscopy, frequency metrology and all applications of lasers requiring high coherence. It is also of fundamental importance, because the Schawlow-Townes laser linewidth limit is of quantum origin. Recently, a theory of stimulated Brillouin laser (SBL) linewidth has been reported. While the SBL linewidth formula exhibits power and optical Q factor dependences that are identical to the Schawlow-Townes formula, a source of noise not present in two-level lasers, phonon occupancy of the Brillouin mechanical mode, is predicted to be the dominant SBL linewidth contribution. Moreover, the quantum-limit of the SBL linewidth is predicted to be twice the Schawlow-Townes limit on account of phonon participation. To help confirm this theory the SBL fundamental linewidth is measured at cryogenic temperatures in a silica microresonator. Its temperature dependence and the SBL linewidth theory are combined to predict the number of thermo-mechanical quanta at three temperatures. The result agrees with the Bose-Einstein phonon occupancy of the microwave-rate Brillouin mode in support of the SBL linewidth theory prediction. 
\end{abstract}

\maketitle

Stimulated Brillouin Scattering (SBS) is a third-order ($\chi^3$) optical nonlinearity that results from the interaction between photons and acoustic phonons in a medium \cite{brillouin1922diffusion,shen1965theory,boyd2003nonlinear,agrawal2007nonlinear}. SBS has practical importance in optical fiber systems \cite{ippen1972stimulated,kobyakov2010stimulated} where it is an important signal impairment mechanism in long-distance transmission systems \cite{Chraplyvy1990} and makes possible all-fiber lasers \cite{stokes1982all} as well as tunable, slow-light generation \cite{okawachi2005tunable}. Power fluctuation resulting from thermal phonons has also been studied in fiber-optic SBS Stokes wave generation \cite{Gaeta1991}. More recently, the SBS process has attracted considerable interest in micro and nanoscale devices \cite{eggleton2013inducing}. Brillouin laser action has been demonstrated in several microcavity resonator systems including silica \cite{Tomes2009,lee2012chemically,li2012characterization,loh2016microrod}, CaF$_{2}$ \cite{grudinin2009brillouin} and silicon \cite{Rakich2017Laser}, and Brillouin amplification has been demonstrated in integrated chalcogenide waveguides \cite{pant2011chip}. In silicon waveguides, the use of confinement to enhance amplification has been studied \cite{kittlaus2016large}. SBS is also a powerful tool for integrated photonics signal processing \cite{Merklein2016,li2013microwave,shin2015coherent}, and it has been applied to realize a chip-based optical gyroscope \cite{Li2017gyro}. Moreover, at radio-frequency rates, the SBS damping rate is low enough in certain systems to enable cavity optomechanical effects \cite{Kippenberg2008} including optomechanical cooling \cite{Bahl2012} and optomechanical-induced transparency \cite{kim2015non}.

This work studies a recent prediction concerning the fundamental linewidth (i.e., non technical noise contribution to linewidth) of the stimulated Brillouin laser (SBL). The formula for SBL linewidth in Hertz (full-width half maximum) \cite{li2012characterization} and the conventional laser Schawlow-Townes linewidth (2-level laser system) \cite{Yariv} are given below,
\begin{equation}
\Delta \nu_{SBS} = \frac{\hbar \omega^3}{4\pi P Q_T Q_E} (n_T + N_T +1),
\end{equation}
\begin{equation}
\Delta \nu_{2-Level} = \frac{\hbar \omega^3}{4\pi P Q_T Q_E} (N_T +{1\over 2})
\end{equation}
where $n_T$ is the number of thermal quanta in the mechanical field at the Brillouin shift frequency, $N_T$ is the number of thermal quanta in the Stokes optical field (negligible at optical frequencies and henceforth ignored), $P$ is the SBL output power, $Q_T$ ($Q_E$) is the total (external) Q-factor, and $\omega$ is the laser frequency. Eq. (1) is valid when the Brillouin gain bandwidth is much broader than the optical cavity linewidth. At very low temperatures where $n_T$ is negligible, the quantum limited SBL linewidth is twice as large as the Schawlow-Townes linewidth on account of phonon participation in the laser process. At finite temperatures $n_T$ is predicted to provide the dominant contribution to the fundamental SBL linewidth. SBL linewidth measurements at room temperature are consistent with this prediction \cite{li2012characterization}. In this study, the phonon contribution to eq. (1) is verified by determination of $n_T$ over a wide range of temperatures using eq. (1) followed by comparison to the Bose-Einstein phonon occupancy at these temperatures.  


\begin{figure*}[htbp]
\centering
\mbox{\includegraphics[width=\linewidth]{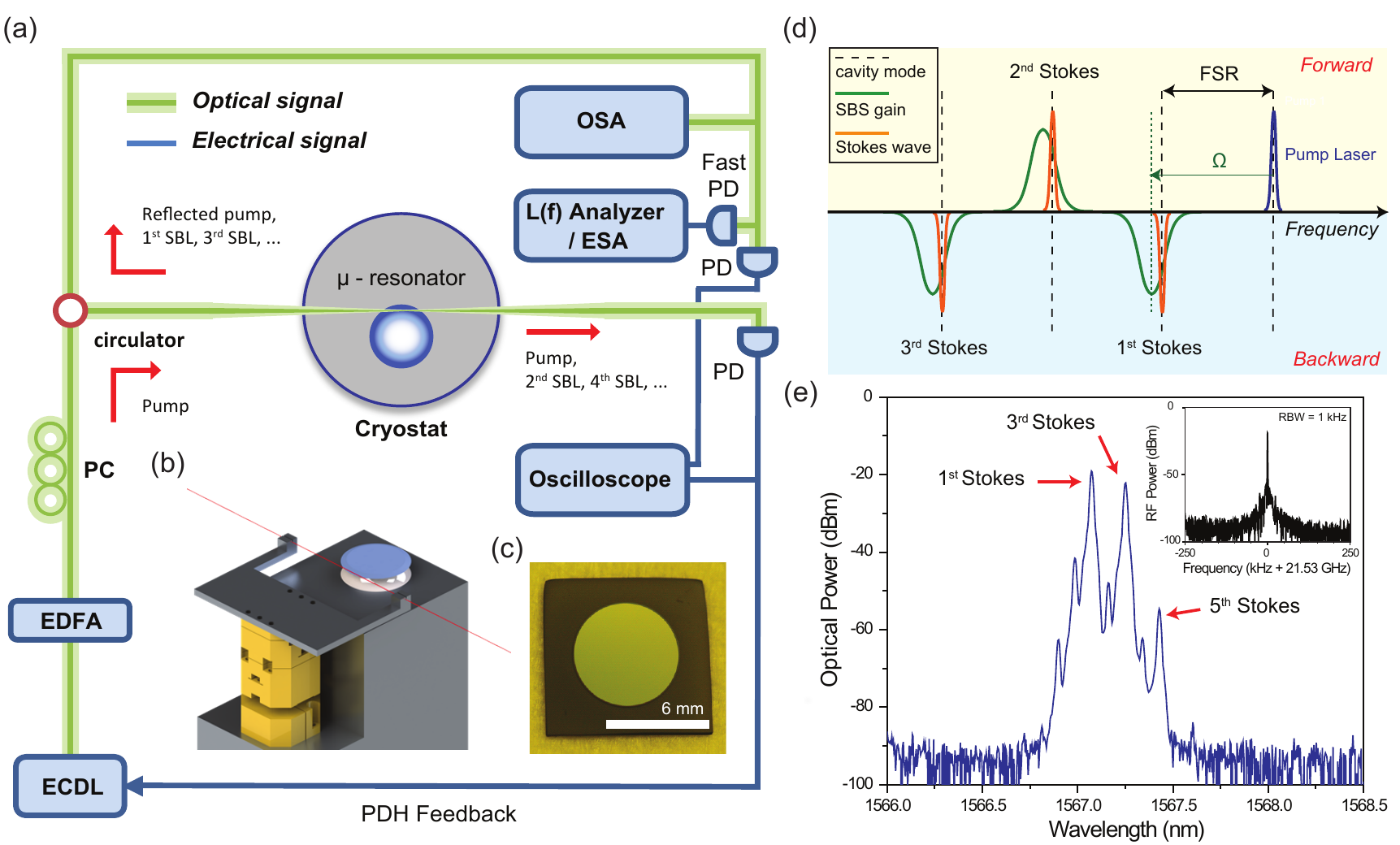}}
\caption{ {\bf Experimental setup and Brillouin laser action} (a) Experimental setup showing external cavity diode laser (ECDL) pump, erbium-doped fiber amplifier (EDFA), polarization control (PC) and circulator coupling to the cryostat. Green lines indicate optical fiber. A fiber taper is used to couple to the microresonator. Pump and even-ordered stimulated Brillouin laser (SBL) waves propagate in the forward direction while odd-ordered SBL waves propagate in the backward direction and are coupled using the circulator. Photodetectors (PD) and an oscillopscope monitor the waves propagating in both directions. A fast photodetector measures the 1\textsuperscript{st}/ 3\textsuperscript{rd} beatnote which is measured using an electrical spectrum analyzer (ESA) and phase noise (L(f)) analyzer. An optical spectrum analyzer (OSA) also measures the backward propagating waves. The pump laser is locked to the microresonator optical resonance using a Pound-Drever-Hall (PDH) feedback loop. (b) Schematic of the optical fiber taper coupling setup inside the cryostat. Optical fiber (red) is glued to an aluminum holder which is fixed on a 3-axis piezoelectric stage. The microresonator is mounted on a copper plate. (c) Top view of the 6 mm wedge disk resonator. (d) Illustration of cascaded Brillouin laser action. Pump and even Stokes orders propagate in the forward direction while odd orders propagate in the backward direction. Green curves represent the Brillouin gain spectra. Brillouin shift frequency ($\Omega$) and free-spectral-range (FSR) are indicated. (e) Optical spectrum measured using the OSA and showing cascaded Brillouin laser action to 5th order.  Inset: typical electrical beatnote spectrum produced by the 1$^{\rm st}$ and 3$^{\rm rd}$ order Stokes laser signals.}
\end{figure*}

Figure 1(a) shows the measurement setup. Pump and signal light are conveyed using fiber optic cable (green lines in figure 1(a)). After passing through an optical circulator, the pump laser passes into the cryostat using a fiber vacuum feedthrough. Inside the cryostat the pump laser power is evanescently coupled to a silica disk microresonator using a fiber taper that is positioned piezoelectrically (Figure 1(b)). Pumping power to the resonator as high as 20 mW was possible. The silica microresonator, shown in figure 1(c), is a wedge design \cite{lee2012chemically}. The resonator diameter was approximately 6 mm to phase match the Brillouin process at cryogenic temperatures (see discussion below). The cryostat is an open-loop continuous-flow unit made by Janis and was cooled to 77 K using liquid nitrogen and to 8 K using liquid helium. 

Brillouin laser action proceeds as diagrammed in figure 1(d) where cascaded lasing is illustrated. Pump light coupled to a resonator mode induces Brillouin gain over a narrow band of frequencies shown in green (typically 20 - 60 MHz wide) that are down-shifted by the Brillouin shift frequency $\Omega / 2 \pi = 2 n V_s / \lambda_P $ where $V_s$ is the sound velocity, $n$ is the refractive index and $\lambda_P$ is the pumping wavelength \cite{li2012characterization}. At room temperature in the silica devices tested here, the Brillouin-shift frequency is 10.8 GHz for optical pumping near 1.55 $\mu$m. When the cavity free-spectral-range (FSR) approximately equals $\Omega$, stimulated Brillouin lasing is possible creating a 1$^{\rm st}$-Stokes laser wave. This wave propagates backward relative to the pump wave on account of the Brillouin phase matching condition, and emerges from the cryostat at the fiber input (figure 1(a)). With increasing pump power, the $1^{\rm st}$-Stokes laser wave will grow in power and ultimately induce laser action on a 2$^{\rm nd}$-Stokes laser wave, which,  by phase matching, propagates in the forward direction. This cascaded laser process is illustrated in figure 1(d) to 3$^{\rm rd}$ order. Phase matching ensures that odd (even) orders propagate backward (forward), and follow fiber-optic paths in figure 1(a) to measurement instruments. Figure 1(e) is a spectrum of cascaded laser action to 5$^{\rm th}$ order measured using the OSA in figure 1(a). Odd orders appear stronger than the even orders (and the pump signal) because the OSA is arranged to detect odd orders (see figure 1(a)). Even-order detection occurs because of weak back-scattering in the optical system.

\begin{figure}
\centering
\includegraphics[width=\linewidth]{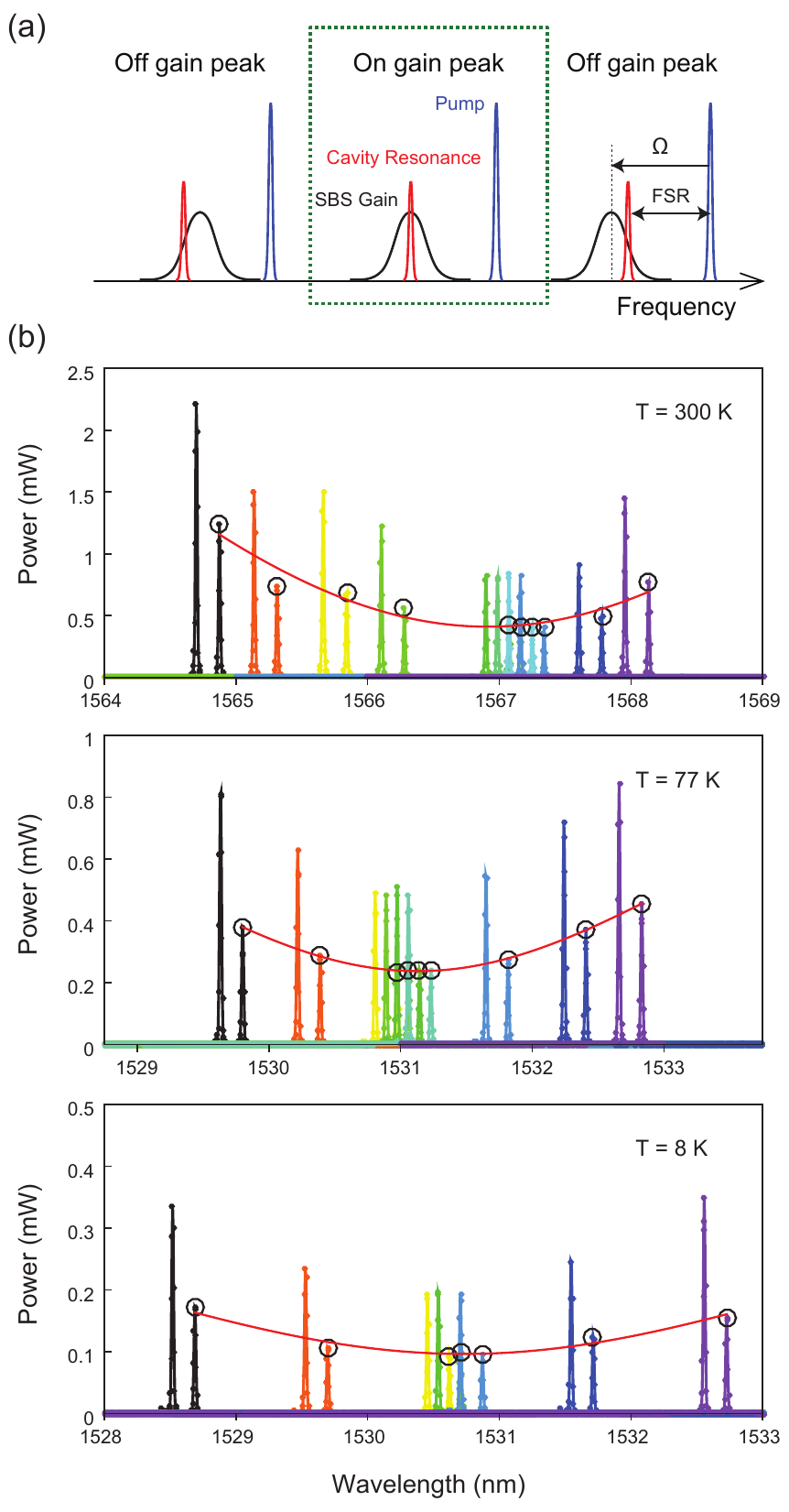}
\caption{ {\bf Aligning the 3$^{\rm rd}$ Stokes wave to the Brillouin gain spectrum maximum} (a) Illustration showing spectral placement of Stokes wave with respect to Brillouin gain spectrum maximum. Variation of the pumping wavelength causes the Brillouin shift frequency ($\Omega$) to vary and thereby scans the Stokes wave across the Brillouin gain peak. (b) Measured spectra of $P_\mathrm{{clamp}}$, the 3\textsuperscript{rd}-order Stokes wave power (indicated by the black circle in the plots). The three panels show measurements performed at T = 300 K, 77 K and 8 K. Each color corresponds to a distinct pumping wavelength. The 1$^{st}$-order Stokes wave also appears in the spectral map as the stronger peak near the 3\textsuperscript{rd}-order Stokes wave. The pump wave is not observable in the linear-scale spectrum as it propagates in the direction opposite to the 1$^{st}$-order and 3\textsuperscript{rd}-order Stokes waves.  $P^{\rm min}_\mathrm{{clamp}}$ is determined (with corresponding pumping wavelength) from the fitted red curve as the minimum power point.}
\end{figure}

The SBL cascade obeys a system of rate equations relating the circulating photon number, $p_n$, of the n$^{th}$-Stokes laser wave to the circulating photon number, $p_{n-1}$, of its preceding (n-1)$^{st}$-pump wave \cite{li2012characterization}.
\begin{equation}
\dot p_n = { g_n } p_{n-1} p_n - {\omega_n \over Q_T} p_n,
\end{equation}
where 
\begin{equation}
g_n = \hbar \omega_n v_g^2 \Gamma \frac{g_B}{V_{eff}} \approx \hbar \omega_n v_g  \Gamma \frac{\Omega}{2\pi} \left( \frac{g_B}{A_{eff}} \right)
\end{equation}
is the Brillouin gain coefficient for the n$^{\rm th}$-Stokes laser wave in Hertz units. Here, $\omega_n$ is the optical frequency of the n$^{\rm th}$ Stokes wave, $v_g$ is the group velocity, $\Gamma$ is the phonon-photon mode overlap factor (defined as the optical mode area, $A_{eff}$, divided by the acousto-optic effective mode area \cite{kobyakov2005design}), $g_B$ is the bulk Brillouin gain coefficient of silica, and $V_{eff}$ is the effective optical mode volume of the n$^{\rm th}$-Stokes laser wave. $g_B / A_{eff}$ is the normalized Brillouin gain coefficient in $W^{-1}m^{-1}$ unit that is typically measured in optical fibers.

The gain coefficient has the spectral profile of the Brillouin gain spectrum (i.e., green curves in figure 1(d)). As pumping to the resonator is increased, a Stokes wave will begin to lase and increase in power until it clamps when the threshold condition for the next Stokes wave in the cascade is reached. This clamped power, $P_{clamp}$, follows directly from the steady-state form of eq. (3), 
\begin{equation}
P_{clamp} = {\omega_{n-1} \over Q_E} {\hbar \omega_{n-1} p_{n-1} } \approx \frac{1}{g} \frac{\hbar \omega^3}{Q_T Q_E}
\end{equation}
where the approximation results from letting $\omega_{n-1} \approx \omega_n$ and in the final result the Stokes order, $n$, is suppressed. At this clamped power, the fundamental linewidth of the Stokes laser mode follows by substitution of eq. (5) into eq. (1), 
\begin{equation}
\Delta \nu_{clamp} =  \frac{g}{4\pi} (n_T+1).
\end{equation}
It is useful to note that the SBL linewidth in the clamped condition is independent of $Q_T$ and $Q_E$. From eq. (5), measurement of $P_{clamp}$, $Q_T$ and $Q_E$ are sufficient to determine $g$. If combined with eq. (6) and measurement of $\Delta \nu_{clamp}$ then $n_T$ can be determined at each operating temperature. 

$g$ depends on the placement of the Stokes wave within the Brillouin gain spectrum (see green spectral curve in figure 1(d)), and the value of $g$ at the spectral maximum (defined as $g_0$) was also determined for comparison with theory.  To determine this maximum, the pump was tuned while recording $P_{clamp}$. This causes the Brillouin shift frequency $\Omega$ to also tune, and therefore to vary the spectral location of the Stokes wave within the Brillouin gain band (see figure 2(a)). $P_{clamp}$ will be minimum (denote as $P^{min}_{clamp}$)  when the Stokes wave is spectrally aligned to the maximum value of $g$, thereby allowing determination of the pumping wavelength corresponding to maximum $g$. At this pumping wavelength, the value $P^{min}_{clamp}$ can be used to determine $g_o$ from eq. (5) when $Q_T$ and $Q_E$ are measured. 

Spectra showing multiple measurements of clamped power for the 3$^{\rm rd}$-Stokes wave at different pumping wavelengths are presented in figure 2(b). The three panels show spectra at T= 300, 77, and 8 K. The 3$^{\rm rd}$-Stokes wave spectral peak at each pumping wavelength is identified by a black circle. At each temperature, the minimum clamped power and corresponding wavelength are determined from the quadratic fit (red curve in figure 2(b)). At this pumping wavelength $g=g_0$ (center case in green box in figure 2(a)). As an aside, the power clamping condition for the 3$^{\rm rd}$-Stokes wave was determined by monitoring the onset of laser action in the 4$^{\rm th}$-Stokes wave. Also, optical losses between the resonator and the OSA were calibrated to determine the clamped power. Table I summarizes the measured minimum clamped powers, $P^{min}_{clamp}$, and their corresponding pumping wavelengths. $Q_T$ and $Q_E$ are also given and were determined by fitting both the linewidth and the transmission minimum of the Stokes mode. Finally, $g_0$, calculated using eq. (5), is compiled in the table I. 

As an aside, the Brillouin gain bandwidth, $\Delta \nu_B$, is also extracted from a quadratic fit of the curves in fig. 2(b) \cite{li2012characterization}. We measure 20 MHz at 300 K, 25 MHz at 77 K, and 35 MHz at 8 K. These linewidths reflect the damping rate of the Brillouin process and have been the focus of theory and experiment in silica optical fiber \cite{le2003study,Rakich2015Helium}. The measured temperature dependence is not consistent with theory and is believed to result from different optical and acoustical mode families participating in the Brillouin process at different temperatures. To partially test this hypothesis, Brillouin linewidths were measured at room temperature by inducing Brillouin laser action on a range of different cavity modes. Linewidths in the range 15 MHz - 45 MHz  were measured  suggesting that damping of the Brillouin process is strongly affected by the spatial structure of the mode. This  could, for example, result from differences in the surface interactions of the various spatial acoustical modes with the wedge resonator dielectric-air interface. It is important to note that this behavior in no way affects the measurement of $n_T$ since  it is the measured value of $g$ and not the theoretical value of $g$ that matters. 


\begin{table}[htbp]
\centering
\caption{\bf Experimental parameters for Brillouin gain ($g_0$) calculation}
\begin{tabular}{cccccc}
\hline
T & $\lambda$ & $Q_{T}$ & $Q_{E}$ & $P_{clamp}^{min}$ & $g_{0}$ \\
( $K$ ) & ( nm )& ( x $10^{6}$ )  & ( x $10^{6}$ )  &  ( mW ) & ( Hz ) \\
\hline
$300$ & 1567.0 & 40  & 50.5  & 0.4  & 0.2272 \\
$77$ & 1531.1 & 82.5  & 91  & 0.2423  & 0.1082 \\
$8$ & 1530.8 & 94  & 103  & 0.0935  & 0.2174\\
\hline
\end{tabular}
\end{table}


To measure the laser linewidth, the beat of the 1$^{\rm st}$ and 3$^{\rm rd}$ Stokes laser waves is detected using a fast photodetector.  An electrical spectrum analyzer trace of this beat is provided as the inset in figure 1(e). As described elsewhere \cite{li2013microwave} the phase noise of this beat signal provides spectral components associated with the fundamental phase noise of the Stokes waves and can be used to infer the linewidth. Moreover, because the 1$^{\rm st}$-Stokes wave has more power than the 3$^{\rm rd}$-Stokes wave (see figure 2(b)) the fundamental linewidth of the 1$^{\rm st}$-Stokes wave is narrower. Accordingly, fundamental phase noise in the beat signal is dominated by the 3$^{\rm rd}$-Stokes wave. Also, microcavity technical frequency noise, while present, is reduced in this measurement because the two Stokes laser waves lase within a single cavity. 

\begin{figure*}[htbp]
\centering
\mbox{\includegraphics[width=\linewidth]{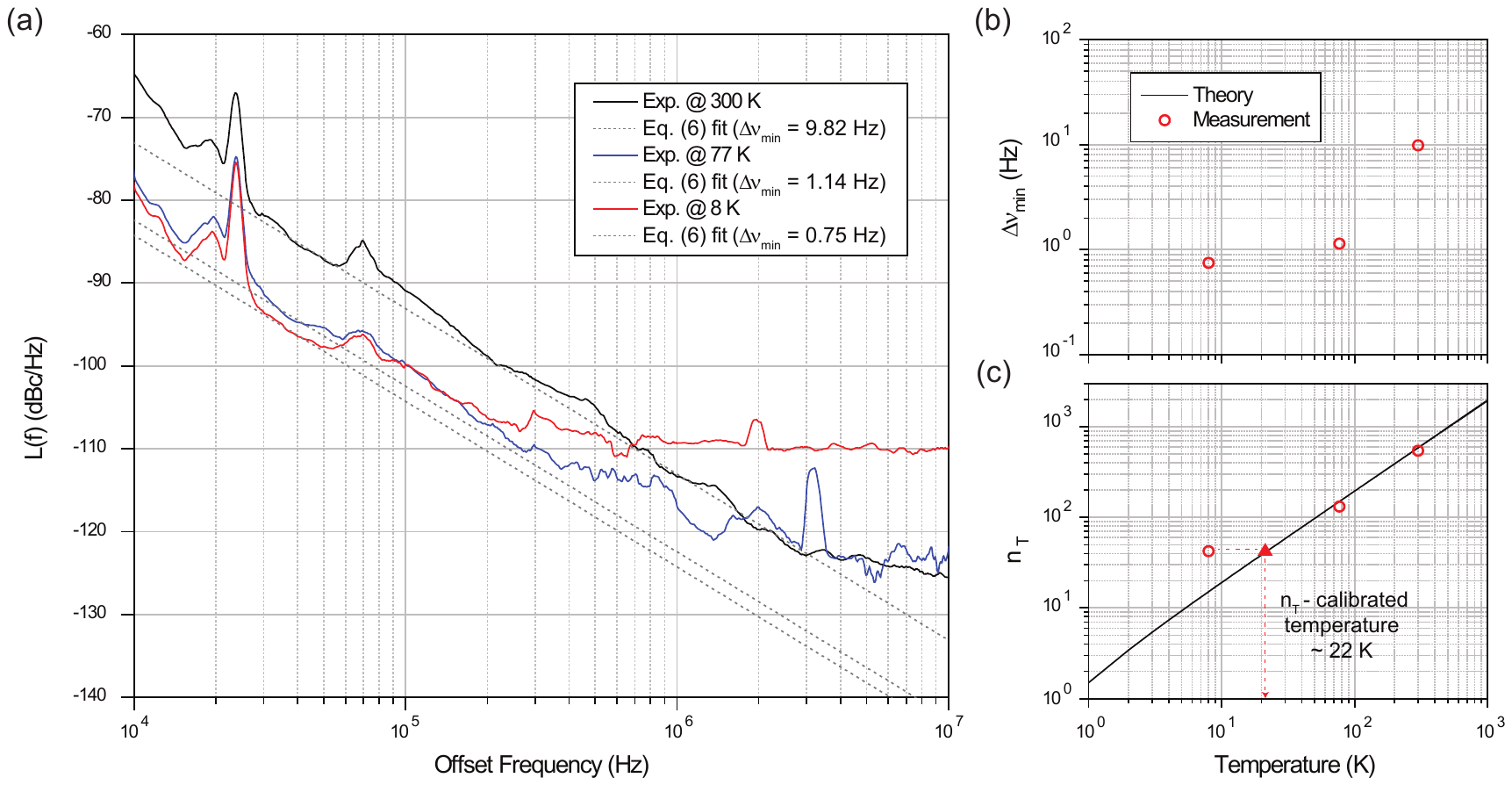}}
\caption{(a) Phase noise spectra of the 1\textsuperscript{st}/3\textsuperscript{rd}-order SBL beatnotes at three temperatures (300 K, 77 K, 8 K). The eq. (7) fit at each temperature is the black dotted line. (b) Red points are minimum SBL linewidth resulting from fitting in panel (a) using eq. (7). (c) Thermal phonon occupancies ($n_{T}$) calculated using eq. (7) from the measured minimum linewidth (panel b plot) and $g_{0}$ (see Table I) are plotted (red points) versus temperature. $n_T$ from the Bose-Einstein occupancy is given as the black line. Triangular point: $n_T$-based temperature calibration to 22 K using Bose-Einstein result.  }
\end{figure*}

The phase noise of 1\textsuperscript{st}/3\textsuperscript{rd}-order SBL beatnote is measured at the $P^{min}_{clamp}$ wavelength determined in figure 2(b). A Rhode-Schwarz FSUP26 phase noise analyzer was used. The measured phase noise spectra at 300 K, 77 K and 8 K are shown in figure 3(a). The theoretical phase noise spectrum for an SBL limited by fundamental noise is given by the expression,
\begin{equation}
L (\nu) = \frac{ \Delta \nu_{min} }{2 \nu^2} = \frac{g_0}{8 \pi \nu^2} (n_T  +1)
\end{equation}
where $\Delta \nu_{min}$ is the fundamental linewidth given by eq. (6) with $g= g_0$ and the second equality in eq. (7) uses eq. (6). The black dashed lines in figure 3(a) give minimum noise-level fits to the measured phase noise spectra using eq. (7). Even with the common-mode noise suppression noted above, there is considerable technical noise coupling to the phase noise spectrum  from the cryogenic system; and fitting is not possible over the entire spectral range. The corresponding $\Delta \nu_{min}$ is plotted in figure 3(b). By using $g_0$ from the table I and the $\Delta \nu_{min}$ data in figure 3(b), eq. (7) provides values for $n_T$ at the three temperatures. These inferred $n_T$ values are plotted versus temperature in figure 3(c). The Bose-Einstein thermal occupancy is also provided for comparison. The discrepancy between the lowest temperature $n_T$ value and the Bose-Einstein value could result from parasitic optical heating or temperature difference between the temperature sensor and the resonator. A calibrated temperature of 22 K is estimated using the Bose-Einstein curve. 

In summary, stimulated Brillouin lasers are unusual because their fundamental linewidth is predicted to be limited by thermo-mechanical quanta of the Brillouin mode. We have confirmed this prediction by determining the thermal phonon occupancy versus temperature using the SBL phase noise. Measurements at 300 K, 77 K and 8 K are in good agreement with the expected Bose-Einstein occupancy. This work provides a possible way to reduce the SBL linewidth for precision measurements. It also lends support to the theoretical prediction that the quantum-limited linewidth of an SBL is strongly influenced by the phonon zero-point motion. It is also worth noting that the rapid increase of $g_{0}$ at lower temperatures in silica indicates that the Brillouin gain bandwidth is expected to be comparable to the optical cavity linewidth below 2 K. In this regime, the system would enter a cavity optomechanical regime \cite{Kippenberg2008} wherein optical damping can exceed mechanical damping. This would require a modification to the SBL linewidth formula. 

This research was supported by the DARPA PULSE program, the Kavli Nanoscience Institute, and by NASA. 

M.-G. Suh and Q.-F. Yang contributed equally to this work. 

\bibliography{bibtex}

\end{document}